\documentclass[a4paper,twocolumn,10pt]{article}
\usepackage[margin=0.6in]{geometry}

\usepackage{authblk}
\RequirePackage{graphicx}
\RequirePackage[colorlinks,citecolor=blue,urlcolor=blue,linkcolor=blue]{hyperref}
\begin{document}

\title{Contribution of inter- and intraband transitions into electron-phonon coupling in metals
}

\author[1]{Nikita Medvedev\footnote{e-mail: nikita.medvedev@fzu.cz} }
\author[2]{Igor Milov }



\affil[1]{Institute of Physics, Czech Academy of Sciences, Na Slovance 2, Prague 8, 18221, Czech Republic}
\affil[1]{Institute of Plasma Physics, Czech Academy of Sciences, Za Slovankou 3, 182 00 Prague 8, Czech Republic}
\affil[2]{Industrial Focus Group XUV Optics, MESA+ Institute for Nanotechnology, University of Twente, Drienerlolaan 5, 7522 NB Enschede, The Netherlands}

\date{Received: date / Accepted: date}

\maketitle

\abstract{We recently developed an approach for calculation of the electron-phonon (electron-ion in a more general case) coupling in materials based on tight-binding molecular dynamics simulations. In the present work we utilize this approach to study partial contributions of inter- and intraband electron scattering events into total electron-phonon coupling in Al, Au, Cu elemental metals and in AlCu alloy. We demonstrate that the interband scattering plays an important role in electron-ion energy exchange process in Al and AlCu, whereas intraband $d-d$ transitions are dominant in Au and Cu. Moreover, inter- and intraband transitions exhibit qualitatively different dependencies on the electron temperature. Our findings should be taken into account for interpretation of experimental results on electron-phonon coupling parameter.}


\section{Introduction}
\label{intro}

Since the advent of powerful femtosecond lasers, the field of material response to irradiation has been developing fast~\cite{Lamb1999}. It is driven by wide range of applications in materials surface and bulk processing and nanostructuring for photonics~\cite{stone2015direct}, catalysis~\cite{nuernberger2012initiation} and biomedicine~\cite{Melissinaki_2011}. Elemental metals and alloys is a class of materials that is widely used in the ultrafast community for its relative simplicity and versatile functionality~\cite{Vorobyev:06,BIZIBANDOKI2013399,doi:10.1021/la2011088}. An ultrafast transfer of the absorbed laser energy from an electronic system of a metal to the lattice is a core process that defines the nature and dynamics of irradiated target evolution. Understanding and quantifying such processes is important to keep advancing the field of ultrafast light-matter interaction.


Most often, response of metals to ultrafast laser pulses is modelled with the two temperature model (TTM) - a set of coupled differential equations for the electronic and atomic/phononic heat conduction and exchange~\cite{M.I.KaganovI.M.Lifshits1956,Anisimov1974}. The latter is controlled with an electron-phonon coupling parameter, which, in a general case, is a function of many variables defining a material transient state, such as electron and ion temperature, density, etc~\cite{Medvedev2020a}. Despite shortcomings of the TTM approach (see e.g. discussions in Refs.~\cite{Huttner1996,Huttner2017}), it remains one of the most widely used models in the community. 

Further refinements of the model are being developed and applied, resulting in multi-temperature approaches, treating different electronic bands and/or different phonon modes separately, each with its own temperature~\cite{Waldecker2016,Hopkins2010}, and hence with different energy exchange among them. Decoupling the contributions into the total coupling from the different electronic bands and interband transitions can help in further development of advanced models that trace different bands separately, such as e.g. in Ref.~\cite{Hopkins2007,Hopkins2009,Ndione2019}. Such models require reliable calculations of various contributions into the coupling parameter.

Here we use the recently developed method of calculating the electron coupling to the ionic motion~\cite{Medvedev2020a} and derive contributions of various inter- (between different electronic bands) and intraband (within one band) electronic transitions to the total coupling parameter in aluminum, copper, gold, and AlCu alloy. We focus on a dependence of these partial couplings on electronic temperature.

\section{Model}
\label{sec:Model}

Electronic coupling to atomic/ionic motion is a process in which an electron transition from one energy level to another occurs, while the energy difference is transferred to or from the atoms. Each atomic displacement induce a change in the Hamiltonian, and correspondingly in its eigenfunctions and eigenstates. These sudden changes from one timestep to another trigger electon transfers between the energy levels~\cite{Tully1990}, known as nonadiabatic coupling between atomic displacement and the electronic wave function. In the solid-state community, it is known as the electron-phonon coupling when the atomic motion is harmonic within an ideal crystal lattice.

We use XTANT-3 method described in Ref.~\cite{Medvedev2020a} to calculate electron-ion coupling parameter of selected materials. We use the term "electron-ion" instead of a more common "electron-phonon" since our model works beyond the harmonic approximation of the atomic system 
(hence it is also capable of calculations of the coupling parameter in disordered matter). The model is based on tight binding molecular dynamics simulations to evaluate the evolution of the Hamiltonian, which is dependent on transient positions of all atoms in the simulation box. Solution of the secular equation provides electron wave functions and eigenstates at each molecular dynamics time step, together with the interatomic forces~\cite{Medvedev2018a}. Knowledge of the transient wave functions allows to calculate the matrix elements of electrons coupling to ionic displacements~\cite{Medvedev2020a}. Using the linear combination of atomic orbitals (LCAO) basis set ($c_{i,\alpha}$) within the tight binding Hamiltonian, a wave function is presented as $\psi_i = \sum_{\alpha} c_{i,\alpha} \phi_{\alpha}$, and the electron transition rate can be written in the following manner~\cite{Medvedev2020a}:

\begin{equation}
\label{eq:1}
w_{i j} = \sum_{\alpha,\beta} w_{ij}^{\alpha \beta} = \frac{4 e}{\hbar \delta t^2} \sum_{\alpha,\beta} \left| c_{i,\alpha}(t) c_{j,\beta}(t_0) S_{\alpha,\beta} \right|^2 
\end{equation}
where $e$ is the electron charge, $\hbar$ is the Planck's constant, $S_{\alpha,\beta}$ is the tight binding overlap matrix, and the wave functions are calculated at two sequential molecular dynamics timesteps $t_0$ and $t = t_0 + \delta t$, where $\delta t$ is the molecular dynamics time step.

The evaluated matrix elements (\ref{eq:1}) are then used in the Boltzmann collision integral to calculate the energy exchange rate between electrons and ions.
The energy exchange rate can then be converted into the electron-ion coupling parameter~\cite{Medvedev2020a}. For each material considered, the molecular dynamics simulations are performed ten times with varied initial conditions (slightly different duration of atomic relaxation prior to the productive simulation run, random atomic velocities corresponding to Maxwellian distribution at the room temperature) and parameters of the electronic temperature increase (increase rate and the maximal aimed temperature), to deliver reliable averaged electron-ion energy exchange rates. In Ref.~\cite{Medvedev2020a} it was demonstrated that this method of calculation provides a good agreement with available experimental data on the electron-phonon coupling in metals at elevated electronic temperatures.

In the present work we present the coupling parameter decomposed into corresponding contributions from the atomic orbitals that form the LCAO basis set ($w_{ij}^{\alpha \beta}$). The employed procedure is analogous to the standard definition of partial density of states (pDOS), or Mulliken atomic charges. It assigns contributions of atomic orbitals into each electronic shell/band, thus dividing the total coupling into partial coupling contributions from electron transitions between the corresponding bands. The interband contributions in our calculations are the sum of direct and inverse transitions: ($\alpha$-$\beta$) is a sum of both, $\alpha \rightarrow \beta$ (terms with $c_{i,\alpha}(t) c_{j,\beta}(t_0)$) and $\beta \rightarrow \alpha$ (terms with $c_{i,\beta}(t) c_{j,\alpha}(t_0)$) transitions. For example, the partial coupling between the states $s$-$p$ is a sum of $s \rightarrow p$ and $p \rightarrow s$, etc.

\begin{figure}
  \includegraphics[width=0.5\textwidth, height=0.3\textwidth, trim=0 10 0 15]{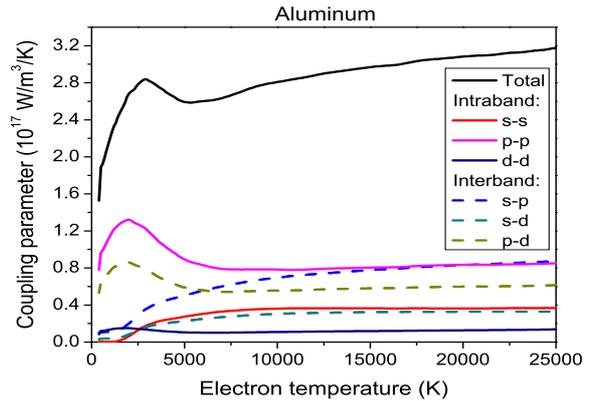}
\caption{Calculated total and partial electron-ion coupling in aluminum.}
\label{fig:Al}
\end{figure}

\begin{figure}
  \includegraphics[width=0.5\textwidth, height=0.3\textwidth, trim=0 10 0 15]{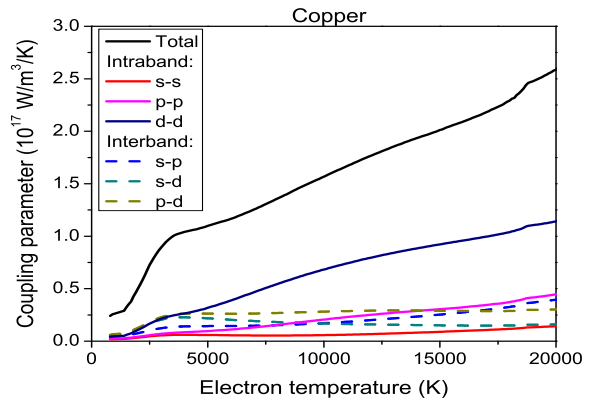}
\caption{Calculated total and partial electron-ion coupling in copper.}
\label{fig:Cu}
\end{figure}
\begin{figure}
  \includegraphics[width=0.5\textwidth, height=0.3\textwidth, trim=0 10 0 15]{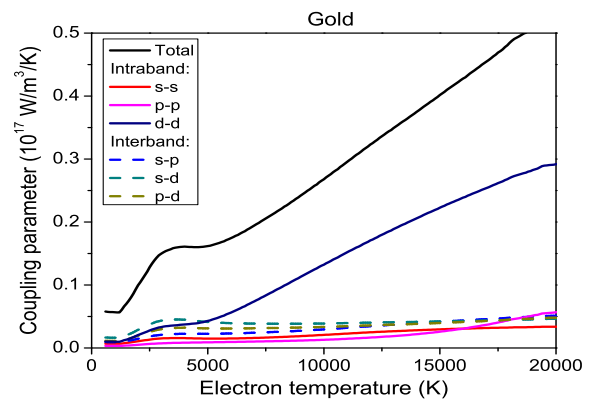}
\caption{Calculated total and partial electron-ion coupling in gold.}
\label{fig:Au}
\end{figure}

The following examples are considered here: elemental metals Al, Au, Cu, and AlCu alloy. For the elemental metals we used the same TB parameterization as in our previous work~\cite{Medvedev2020a} (NRL TB parameterization~\cite{Papaconstantopoulos2003}), whereas for AlCu alloy we used DFTB matsci-0-3 parameterization~\cite{Frenzel2009}. It is also, like NRL, based on sp$^3$d$^5$ LCAO basis set. We assumed AlCu to be in a simple cubic lattice structure, and used a supercell with 250 atoms. We used a 1 fs timestep in the molecular dynamics part of our model.

\section{Results}
\label{sec:Results}

Figure~\ref{fig:Al} shows the coupling parameter in aluminum. The partial couplings attributed to various electronic transitions have qualitatively different dependencies on the electronic temperature. At low electronic temperatures, the main contribution to the total coupling is from the transitions between $p$-$p$ and $p$-$d$ shells, whereas with the increase of the electronic temperature, contributions from interband $s$-$p$ and $s$-$d$ transitions, as well as from the intraband $s$-$s$ transition, start to play a larger role. The $d$-$d$ transitions make only a minor contribution into the total coupling parameter.

\begin{figure*}
 \centering \includegraphics[width=0.6\textwidth, height=0.6\textwidth, trim=20 30 120 20]{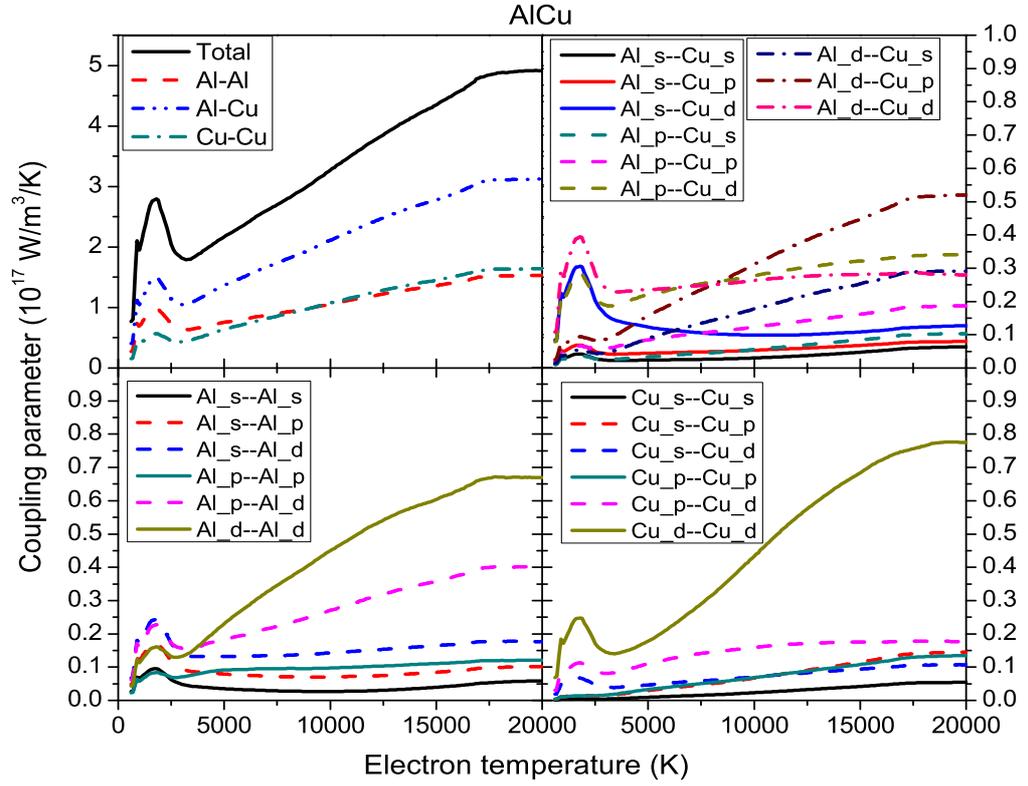}
\caption{Calculated electron-ion coupling in AlCu alloy. Top left panel: total coupling and coupling decomposed into transitions between bands formed by Al and Cu atoms. Top right panel: Interatomic interband contributions into coupling. Bottom left panel: Al inter- and intraband contributions. Bottom right panel: Cu inter- and intraband contributions.}
\label{fig:AlCu}
\end{figure*}

In contrast to aluminum, the main contribution to the increasing total coupling parameter in copper (Figure~\ref{fig:Cu}) and gold (Figure~\ref{fig:Au}) comes from $d$-$d$ transitions. Partial couplings attributed to other transitions exhibit a weak dependence on the electronic temperature and have smaller contributions to the total coupling parameter. In gold, $d$-band containing 10 electrons is located at $\sim 2$ eV below the Fermi level~\cite{Papaconstantopoulos2015}, thus at low electronic temperatures this band does not participate in electron-ion coupling. With increase of the electron temperature, more and more electrons from the $d$-band are involved in the coupling, which increases $d$-$d$ partial coupling, making it dominant at electronic temperatures above $\sim 5000$ K. A similar concept for copper was discussed in Ref.~\cite{Hopkins2008}. In contrast, the partial DOS in aluminum does not have such a strong localization~\cite{Papaconstantopoulos2015} and all the bands contribute into the electron-ion coupling, especially at elevated electronic temperatures. 

In AlCu alloy, we decomposed the contributions into bands formed by shells of different atoms (Al and Cu, and interatomic contributions), and into transitions between various shells in those atoms, which is shown in different panels in Figure~\ref{fig:AlCu}. There, the dominant contributions are from interatomic transitions between the bands formed by Al and Cu atoms (top left panel). Contributions of interband transitions are dominant, however with noticeable intraband transitions within $d$ bands of Al and of Cu (other panels in Figure~\ref{fig:AlCu}). Note again that at different electronic temperatures, different bands have dominant contributions. Scattering within $d$ bands of Al and of Cu atoms in AlCu alloy, as well as interband transitions that involve $d$ bands, have a dominant effect. At low electronic temperatures, the main peak is formed by the transitions between the $d$ band of Cu and other bands of Al. With increase of the electronic temperature, the scattering between $d$-band electrons of Al and other bands increases (as well as $d$-$d$ Cu transitions).

\section{Discussion}
\label{sec:Discussion}

In all calculated materials, most prominently in aluminum and AlCu alloy, interband electron transitions play a significant role in electron-ion energy exchange. Such observation indicates that models accounting only for intraband coupling, even if band-resolved (such as, e.g., in Refs.~\cite{Petrov2013,Ndione2019} where $s$ or $sp$ and $d$ bands were considered to be coupling to phonons separately), may overlook important contributions. Even in gold and copper, albeit the $d$-$d$ contribution is dominant at high electronic temperatures, it is still necessary to account for interband transitions. The importance of interband transitions for electron-phonon coupling was also discussed in a series of works by Hopkins et al.~\cite{Hopkins2010,Hopkins2007,Hopkins2009,Hopkins2008}, albeit within the Eliashberg approximation~\cite{Lin2008} which overestimates the coupling parameter at non-zero electronic temperatures~\cite{Medvedev2020a}.

The presented calculations assumed thermal equilibrium within the electronic system - the electron distribution function always follows the Fermi-Dirac distribution, with the same temperature and chemical potential in all electronic bands. In experiments, it may not always be the case. When electrons are predominantly excited from one band into another one~\cite{Ndione2019}, for example due to particular tuning of the optical photon energy, particular electron transitions may be enhanced or inhibited. For example, one may expect that excitation of electrons only from the $s$-band of gold by photon energy below $2$ eV (without reaching $d$-band electrons), may suppress the electron coupling due to a lack of $d$-$d$ transitions, which are the dominant contribution into the increase of the coupling with the rise of the electronic temperature, see Figure~\ref{fig:Au}. In such a nonequilibrium case, with only $s$ but not $d$-electrons excited, the contribution into the coupling will be smaller than in case when $d$-electrons are contributing (such as in the equilibrium case). It was suggested that this kind of nonequilibrium within the electronic system may last for significantly long times in gold~\cite{Ndione2019}. It was demonstrated experimentally for nickel that irradiation with photon energies exciting $d$-band induces significantly different electron-phonon coupling compared to irradiation with photon energies insufficient to excite $d$-band electrons~\cite{Hopkins2007}.

This fact has large implications for experiments with ultrafast optical laser irradiation. Excitation of electrons into nonequilibrium state may reduce the observable coupling parameter with respect to the situation when electrons are in thermal equilibrium (which corresponds to the total coupling in Figures~\ref{fig:Cu} and \ref{fig:Au} below). This may be the reason why experimental works on ultrafast irradiation of gold often report constant or nearly constant coupling parameter at different electronic temperatures~\cite{Ng2012,Chen2013,Mo2018a}.

\section{Conclusions}
\label{sec:Conclusions}

We calculated partial electron-ion coupling parameters in Al, Cu, Au and AlCu alloy. Contributions of inter- and intraband electronic transitions are presented, which may be useful for models beyond the two temperature approximation. We found that the intraband contribution in all materials, most pronouncedly in copper and gold, behave qualitatively different from interband contributions with increase of the electronic temperature. In particular, $d$-$d$ electron transitions in gold and copper play the dominant role at electron temperature above $\sim 5000$ K. This fact have implications for interpretation of ultrafast light-matter interaction experiments, in which electronic system may be out of equilibrium due to a particular choice of incident photon energy.

\section{acknowledgements}
The authors thank B. Rethfeld, P. Ndione and A. Ng for motivating discussions. 
This work benefited from networking activities carried out within the EU funded COST Action CA17126 (TUMIEE) and represents a contribution to it.
The authors gratefully acknowledge financial support from the Czech Ministry of Education, Youth and Sports (Grants No. LTT17015, No. EF16\_013/0001552, and No. LM2015083). I. Milov gratefully acknowledges support from the Industrial Focus Group XUV Optics of the MESA+ Institute for Nanotechnology of the University of Twente; the industrial partners ASML, Carl Zeiss SMT GmbH, and Malvern Panalytical, and the Netherlands Organisation for Scientific Research (NWO).

\nopagebreak
\onecolumn
\nopagebreak
\bibliographystyle{spphys}       
\nopagebreak
\bibliography{My_Collection_2}   
\nopagebreak

\end{document}